\documentclass{article}

\usepackage[english]{babel}

\usepackage[letterpaper,top=2cm,bottom=2cm,left=3cm,right=3cm,marginparwidth=1.75cm]{geometry}

\usepackage[colorlinks=true, allcolors=blue]{hyperref}
\usepackage{amsmath} 
\usepackage{amssymb} 
\usepackage{authblk}
\usepackage{braket}
\usepackage{csquotes}
\usepackage{graphicx} 
\usepackage{hyperref} 
\usepackage{xurl} 
\usepackage{booktabs} 
\usepackage{enumitem} 
  \usepackage[
    backend=biber,
    style=numeric,
  ]{biblatex}

\newcommand{\keywords}[1]{%
  \vspace{1em}\noindent\textbf{Keywords:} #1
}

\newcommand{\qbit}{\text{qubit}}
\newcommand{\qbits}{\text{qubits}}
\newcommand{\singleqbit}{\text{single-qubit}}
\newcommand{\multiqbit}{\text{multi-qubit}}
\newcommand{\qreg}{\text{quantum register}}
\newcommand{\qregs}{\text{quantum registers}}

\newcommand{\BlochSphere}{\text{Bloch Sphere}}
\newcommand{\Stateogram}{\text{state-o-gram}}
\newcommand{\VENUS}{\text{VENUS}}

\title{\Stateogram{} -- A Novel 2D Visualization for Quantum States}

\author{Fritz Schinkel}
\affil{\textit{\small Fujitsu Germany GmbH, Mies-van-der-Rohe-Straße 8, 80807 Munich, Germany}}

\begin{document}
\maketitle

\begin{abstract}
Quantum computing is rapidly gaining popularity, necessitating intuitive visualization tools for complex quantum states. While the Bloch Sphere effectively visualizes \singleqbit{} states, it fundamentally lacks scalability for \multiqbit{} systems. Existing \multiqbit{} visualization attempts, such as \VENUS{}, have shown promise but often face limitations in scalability beyond a few \qbits{}. This paper introduces \Stateogram{}, a novel 2D visualization approach designed to intuitively represent quantum states for an arbitrary number of \qbits{}. \Stateogram{} effectively visualizes probability amplitudes and phase angles in a unified 2D framework, addressing the limitations of prior art. We detail its design principles, visual elements, and application to \multiqbit{} systems, aiming to provide a scalable and intuitive tool for quantum state analysis. We evaluate the applicability by visualizing the states throughout the Deutsch-Josza algorithm.
\end{abstract}


\section{Introduction}
Quantum computing has seen rapid advancements in recent years, driven by its potentially exponential computational power and potential applications across diverse fields such as optimization, machine learning, and cryptography. At the heart of this technology lie the unique properties of quantum bits (\qbits{}), namely superposition and entanglement. Understanding and manipulating these quantum phenomena are crucial for designing and debugging quantum algorithms. However, the abstract and high-dimensional nature of quantum states often makes them challenging to grasp intuitively.

Quantum state visualization has emerged as a vital tool to address this challenge. The most widely known single-\qbit{} visualization tool, the \BlochSphere{} \cite{Bloch1946, Arecchi1972}, intuitively represents a \qbit{}'s state as a point on a 3D sphere. While effective for single \qbits{} and widely adopted in quantum computing education and debugging (e.g., in IBM Qiskit \cite{ibm14}), its applicability is limited to single-\qbit{} systems. It fundamentally fails to effectively represent superposition in multi-\qbit{} systems, which becomes an increasingly critical issue as quantum computing advances and multi-\qbit{} systems play a central role.

Existing research has made several attempts to visualize multi-\qbit{} systems. For instance, \VENUS{} \cite{Ruan2023} proposed a novel visualization approach using 2D geometric shapes to represent single- and two-\qbit{} quantum states. It successfully visualizes the correlation between amplitudes and probability distributions for up to two \qbits{}. However, existing methods like \VENUS{} still face challenges for more than two \qbits{}. Specifically, a comprehensive and intuitive 2D visualization for arbitrary multiple \qbits{} remains an open problem.

This paper aims to bridge this gap by proposing a new 2D quantum state visualization approach called \textbf{\Stateogram{}}. \Stateogram{} is designed to be scalable not only for single \qbits{} but also for arbitrary multiple \qbits{}, visualizing quantum state probability amplitudes and phase angles in a unified and intuitive 2D representation. Our approach seeks to effectively manage the complexity of the state space, enabling users to easily comprehend subtle changes and interactions within quantum states. In this paper, we detail the design principles, visual elements, and application of \Stateogram{} to multi-\qbit{} systems. Furthermore, we will evaluate its effectiveness and practicality by applying it to the Deutsch-Josza algorithm \cite{Deutsch1992}.

\section{Background and Prior Art}
Quantum state visualization has been a significant area of research for many years, crucial for understanding and developing quantum computing. This section provides an overview of key prior art, highlighting their strengths and limitations, particularly concerning the challenges in multi-\qbit{} systems.

\subsection{Fundamentals of Quantum Computing}
Quantum computing leverages unique quantum mechanical phenomena. A \qbit{} is the basic unit of quantum information, which unlike classical bits, can exist in a superposition of states, representing both 0 and 1 simultaneously. When multiple \qbits{} are linked in a way that their states are correlated regardless of their physical separation, they are said to be entangled. Mathematically, a quantum state of a \multiqbit{} system of $n$ \qbits{} can be described by a state vector in a $2^n$-dimensional Hilbert space. The computational basis is an orthonormal basis of that space. The basis vectors correspond to the numerical values given by the binary interpretation of its \qbits{}. 

\subsection{Existing Quantum State Visualization Methods}
To address the challenge of understanding quantum states, various visualization methods have been developed.

\subsubsection{Bloch Sphere: The Standard for Single-Qubit Visualization}
The \BlochSphere{} \cite{Bloch1946} is the most widely adopted visualization tool, representing a single \qbit{}'s state as a point on a 3D sphere. It intuitively shows the superposition of a \qbit{} (a linear combination of $|0\rangle$ and $|1\rangle$) by the position of a point on the sphere's surface. Due to its simplicity and intuitiveness, it is widely used in quantum computing education and for debugging single-\qbit{} operations. It is integrated into many quantum computing toolkits, such as IBM Qiskit \cite{ibm14}.

However, the \BlochSphere{} has fundamental limitations. Most importantly, it \textbf{does not scale to multi-\qbit{} systems}. States of two or more \qbits{} cannot be represented on a 3D \BlochSphere{} \cite{BSR21}. In particular, multi-\qbit{} specific phenomena like quantum entanglement cannot be visualized at all using the \BlochSphere{}. Additionally, its lack of intuitive display for measurement probabilities and the inherent occlusion problems of 3D representations can hinder precise state analysis \cite{Wil21}.

\subsubsection{Attempts to Extend to Multi-Qubit Systems}
To overcome the limitations of the \BlochSphere{}, many researchers have proposed visualization approaches for multi-\qbit{} systems. These attempts can be broadly categorized as follows:

\begin{itemize}[leftmargin=*]
    \item \textbf{Extensions of \BlochSphere{}:} Some research attempts to extend the \BlochSphere{} concept to multi-\qbit{} systems \cite{AJMK09, HD04}. For example, Q-Sphere \cite{ibm14} represents multiple \qbit{} states on a single sphere, but still faces challenges in fully visualizing entanglement and scalability as the number of \qbits{} increases. These extensions often rely on projections of higher-dimensional spaces or complex geometric structures, making intuitive understanding difficult.

    \item \textbf{2D Geometric Representations:} To circumvent the 3D constraint of the \BlochSphere{}, 2D geometric shape-based visualization approaches have also been developed.
    \begin{itemize}[leftmargin=1.5em]
        \item \textbf{\VENUS{} \cite{Ruan2023}:} This visualization method uses 2D geometric shapes (right triangles and semicircles) to represent single- and two-\qbit{} quantum states. \VENUS{} showed promise by visually correlating amplitudes and probability distributions, particularly in representing two-\qbit{} entanglement. However, \VENUS{} also identifies scalability beyond two \qbits{} as future work.
        \item \textbf{Fractal Representations \cite{GI12}:} Attempts have been made to represent multi-\qbit{} systems using fractal structures, but these also have limitations in intuitively conveying the specific numerical meaning of the state.
        \item \textbf{Density Matrix Mappings \cite{CMM17}:} Methods have been proposed to map the density matrix of \qbits{} onto the vertices of a triangle. However, these are primarily mathematical representations and can be difficult for non-experts to understand.
    \end{itemize}

    \item \textbf{Circuit-Based Visualizations:} There are also visualization tools that focus on the structure of quantum circuits and gate operations \cite{PRZ17, ZW18, RWJ2022, TPCW17, Qui14}. While these tools are helpful for understanding the execution flow of quantum programs, they do not delve into visualizing the detailed internal structure of quantum states themselves (amplitudes, probabilities, phase angles).
\end{itemize}

\subsubsection{Strengths and Limitations of Each Method, Especially Challenges in Multi-Qubit Visualization}
While existing visualization methods have demonstrated utility in specific aspects, they commonly face the following challenges:

\begin{enumerate}[leftmargin=*]
    \item \textbf{Lack of Scalability:} Most methods cannot cope with the exponential complexity of the state space as the number of \qbits{} increases. This makes it difficult to comprehensively and intuitively visualize multi-\qbit{} states.
    \item \textbf{Information Overload and Lack of Intuition:} Attempts to directly visualize complex mathematical representations can lead to information overload, making intuitive understanding difficult for non-experts.
    \item \textbf{Tracking Dynamic State Changes:} The ability to display dynamic changes in quantum states due to quantum gate operations in a real-time and understandable way is often lacking.
\end{enumerate}

These challenges highlight the need for more advanced and scalable quantum state visualization tools for quantum computing education, algorithm development, and debugging. The proposed \Stateogram{} in this research aims to address these limitations of prior art by providing an intuitive 2D representation for arbitrary  \multiqbit{} systems, encompassing probability amplitudes and phase angles in a unified view.

\section{\Stateogram{} -- 2D Quantum State Visualization}
The idea and the name of the \Stateogram{} are inspired by the histogram of basis state probabilities. The histogram represents every vector of the computational basis as one bar. The height represents the probability to measure the respective basis state. The number of bars is the dimension of the state space, so for $n$ \qbits{} we get $2^n$ bars. So the number of bars needed for the histogram grows exponentially. The heights of the bars represent the probabilities of all possible measurements and therefore sum up to $1$. So the diagram becomes very sparse, or the bar heights become infinitesimally small, rendering the visualization unreadable. Crucially the histogram misses the phase angle information, so it does not represent the complete information of a \qbit{} or a \qreg{}.

For the \Stateogram{} we propose a stacked bar chart. We distinguish the basis states by color, e.g. in the numerical order of the computational basis from blue to red. And we use the $x$-axis of the bar chart for the angle of the complex phase of the respective basis vector. So the \Stateogram{} can keep its size with a height of $100\%$ and a width of $2 \pi$. However, the chart can also become fragmented due to very narrow stacked bars. Nevertheless, it will be demonstrated that often not individual bars, but rather the collective arrangement and positioning of bars within the \Stateogram{} convey information about the effects of specific gate operations and the evolution of the state under a particular gate operation or a complete algorithm. The representation of the angle in the \Stateogram{} shows the effects of interference and a sequence of \Stateogram{} charts gives a complete picture of a state superposition at each step of a circuit execution.

\subsection{\Stateogram{} for \singleqbit{} Systems}
\begin{figure}
\centering
\includegraphics[width=0.6\linewidth]{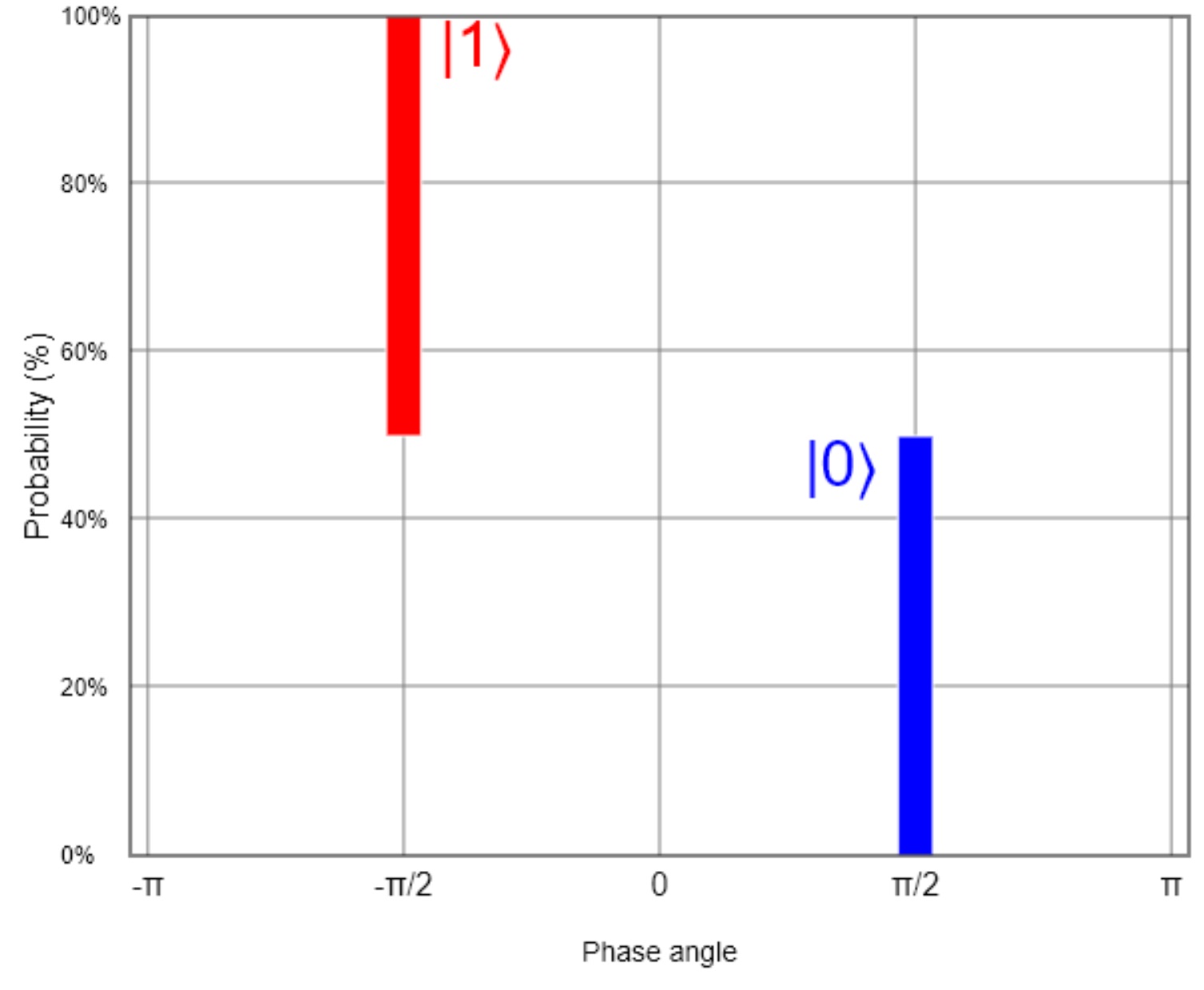}
\caption{\label{fig:single}\Stateogram{} for single \qbit{}.}
\end{figure}

We start with the smallest possible case for the state visualization. Figure \ref{fig:single} shows a \Stateogram{} for a \singleqbit{}. We see the display for a superposition state:
$$
\psi = \frac{i}{\sqrt{2}} \ket{0} - \frac{i}{\sqrt{2}} \ket{1} .
$$
The bar for $\ket{0}$ is shown in blue with the label at the top. The angle of its phase is $\frac{\pi}{2}$ since $i=e^\frac{\pi i}{2}$. So the bar is positioned on the $x$-axis, which is running from $-\pi$ to $\pi$, at $\frac{\pi}{2}$. The probability for measuring $\ket{0}$ is $\frac{i}{\sqrt{2}} \overline{\frac{i}{\sqrt{2}}} = \frac{1}{2}$, therefore the bar has a height of $50\%$. The red bar is drawn analogously at$-\frac{\pi}{2}$ as a stacked bar, so it ranges from $50\%$ to $100\%$. Note that we don't use the magnitude of the amplitude but the square of the magnitude; so the stacked bars always sum up to $100\%$ and reach the upper edge of the diagram.  

\subsection{\Stateogram{} for Multi-Qubit Systems}
\begin{figure}
\centering
\includegraphics[width=0.6\linewidth,page=4]{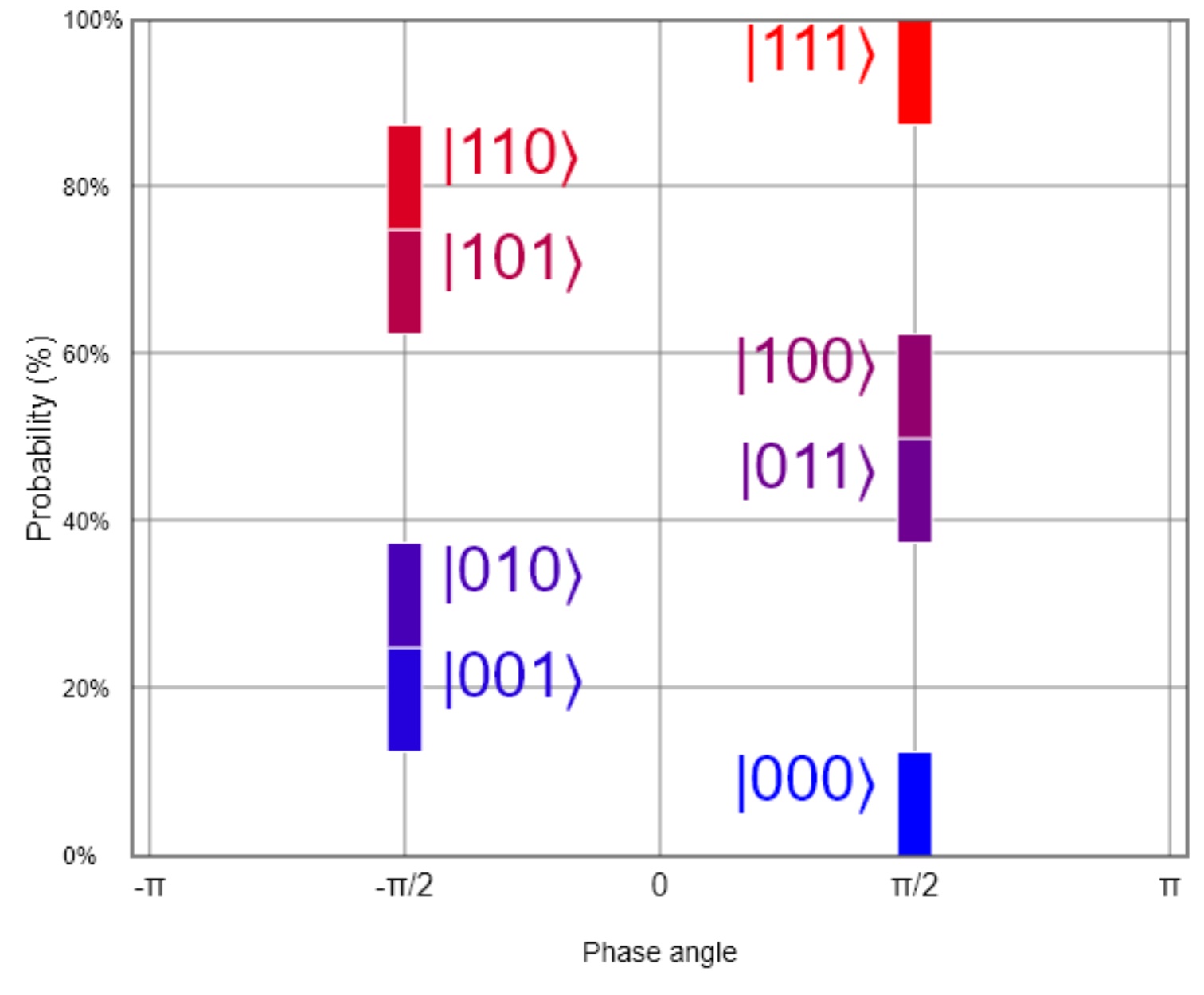}
\caption{\label{fig:multi}\Stateogram{} for 3-\qbit{} register with homogeneous probability amplitude amounts.}
\end{figure}

Next we outline the construction for the general case of a \multiqbit{} register. Figure \ref{fig:multi} shows the result for a $3$-bit register which is in a superposition state of  all basis states contributing with equal magnitude of amplitudes, resulting in equal measurement probabilities:

$$
\psi = \frac{i}{\sqrt{2^3}} \ket{000} - \frac{i}{\sqrt{2^3}} \ket{001} - \frac{i}{\sqrt{2^3}} \ket{010} + \frac{i}{\sqrt{2^3}} \ket{011}
+ \frac{i}{\sqrt{2^3}} \ket{100} - \frac{i}{\sqrt{2^3}} \ket{101} - \frac{i}{\sqrt{2^3}} \ket{110} + \frac{i}{\sqrt{2^3}} \ket{111} .
$$

The bars are colored, shading from blue to red. Each non vanishing basis state with a phase different from $0$ is represented by a bar with a label positioned at the top. It is positioned at the phase angle and has the height of the square of the magnitude of the respective probability amplitude; so it displays the measurement probability of the respective basis state. All bars are stacked, so they start on the $y$-axis with the accumulated sum of probabilities for all basis states with lower index in the computational basis than the considered basis states. Each non vanishing basis state has its own vertical space, thereby facilitating clear labeling without visual overlaps.

\begin{figure}
\centering
\includegraphics[width=0.6\linewidth]{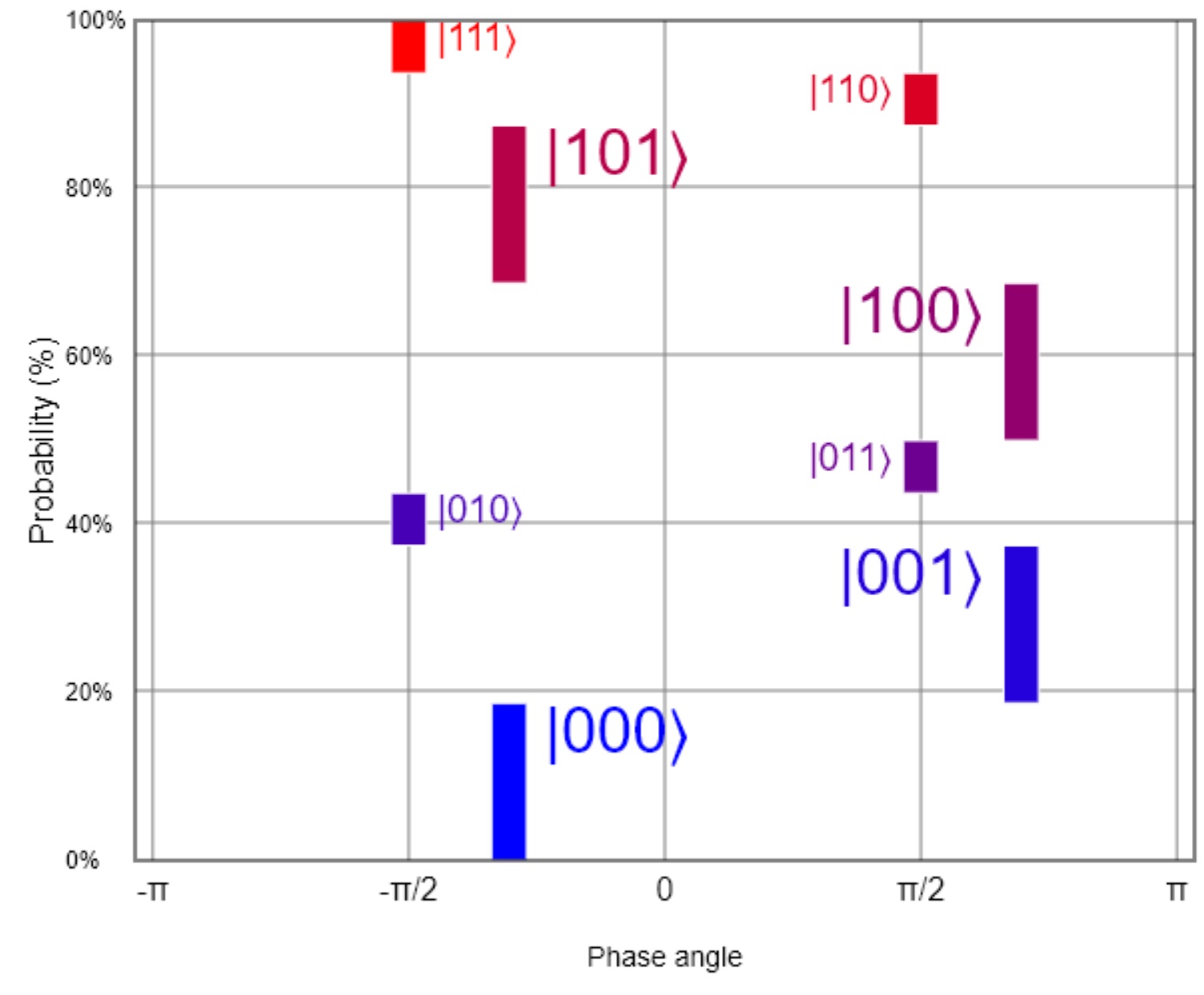}
\caption{\label{fig:multi2}\Stateogram{} for 3-\qbit{} register with heterogeneous probability amplitude amounts.}
\end{figure}

This concept can be applied to any state of an arbitrary register with unequal phase amounts and all angles between  $-\pi$ and $\pi$. An example is shown in figure \ref{fig:multi2}. 

\begin{figure}
\centering
\includegraphics[width=1.0\linewidth]{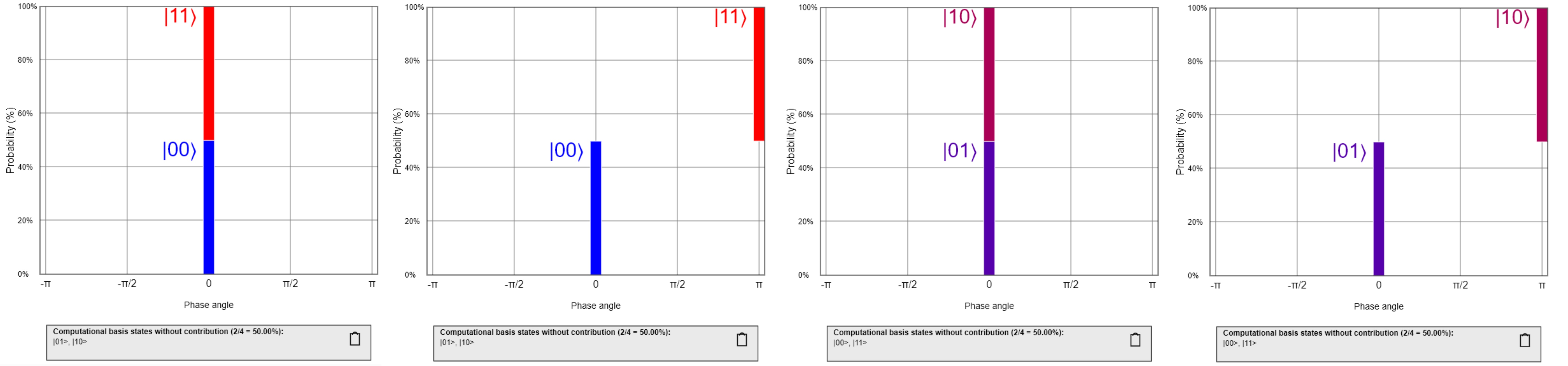}
\caption{\label{fig:bell}\Stateogram{} for the 4 Bell states with vanishing basis states listed in gray box below.}
\end{figure}

Often the state does not contain all basis states. In that case the colors are chosen according to the index of the non-zero vectors of the computational basis. The vanishing basis states with an amplitude of zero are simply listed to inform the reader about their absence from the visualization. As an example we see the Bell states in figure \ref{fig:bell}.

\section{Examples}
In this chapter we show visualization examples. We start with some simple gate operations and their effects on a \multiqbit{} register. We will then systematically analyze the Deutsch-Josza algorithm to elucidate its mechanism for distinguishing between constant and balanced oracles.

\subsection{Hadamard Gate}
\begin{figure}
\centering
\includegraphics[width=1.0\linewidth]{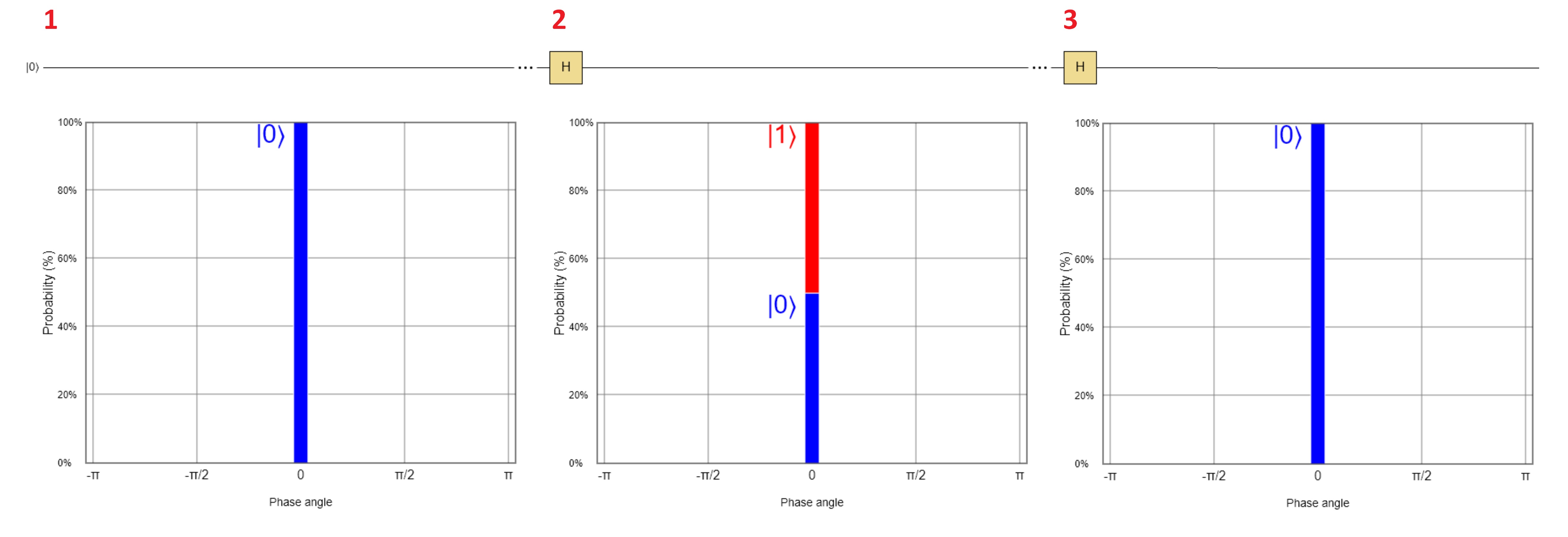}
\caption{\label{fig:Hadamard 1q}Hadamard for 1-\qbit{} register.}
\end{figure}

We begin with a \singleqbit{} initialize to $\ket{0}$. Figure \ref{fig:Hadamard 1q} shows this initial state (1). Next the superposition state $\frac{1}{\sqrt{2}}\ket{0}+\frac{1}{\sqrt{2}}\ket{1}$ after applying a Hadamard gate is displayed (2). Finally we see the  $\ket{0}$ state once again after execution of the Hadamard gate a second time (3).

\begin{figure}
\centering
\fbox{
    \vbox{
        \offinterlineskip 
        \hbox{\includegraphics[width=0.97\linewidth]{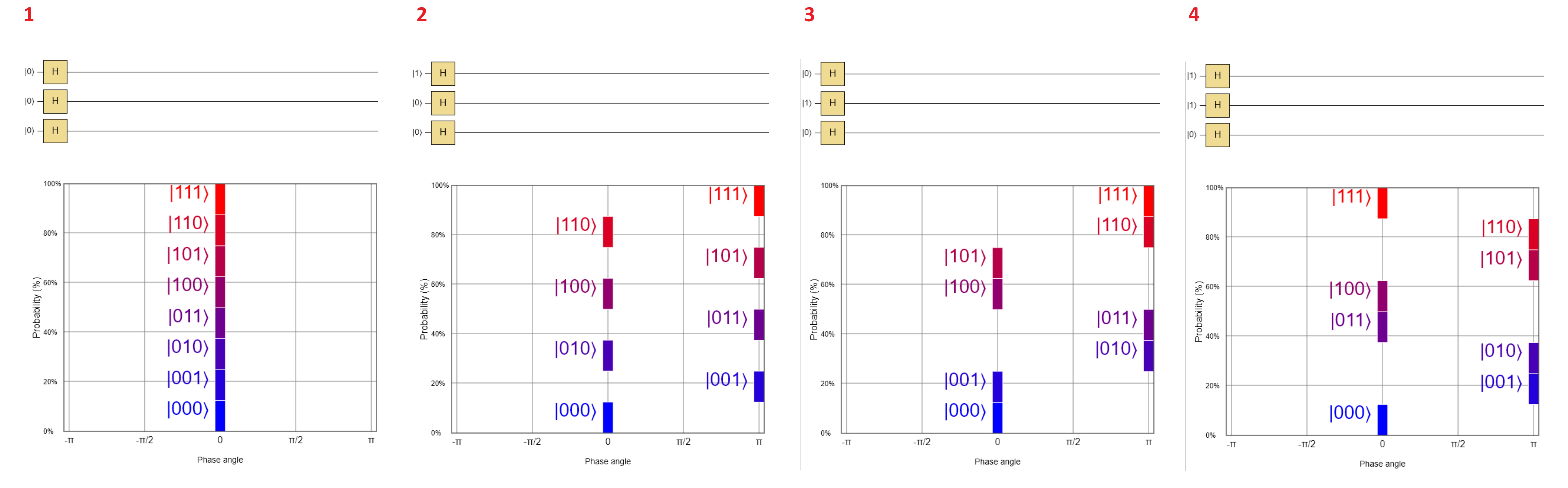}}
        \hbox{\includegraphics[width=0.97\linewidth]{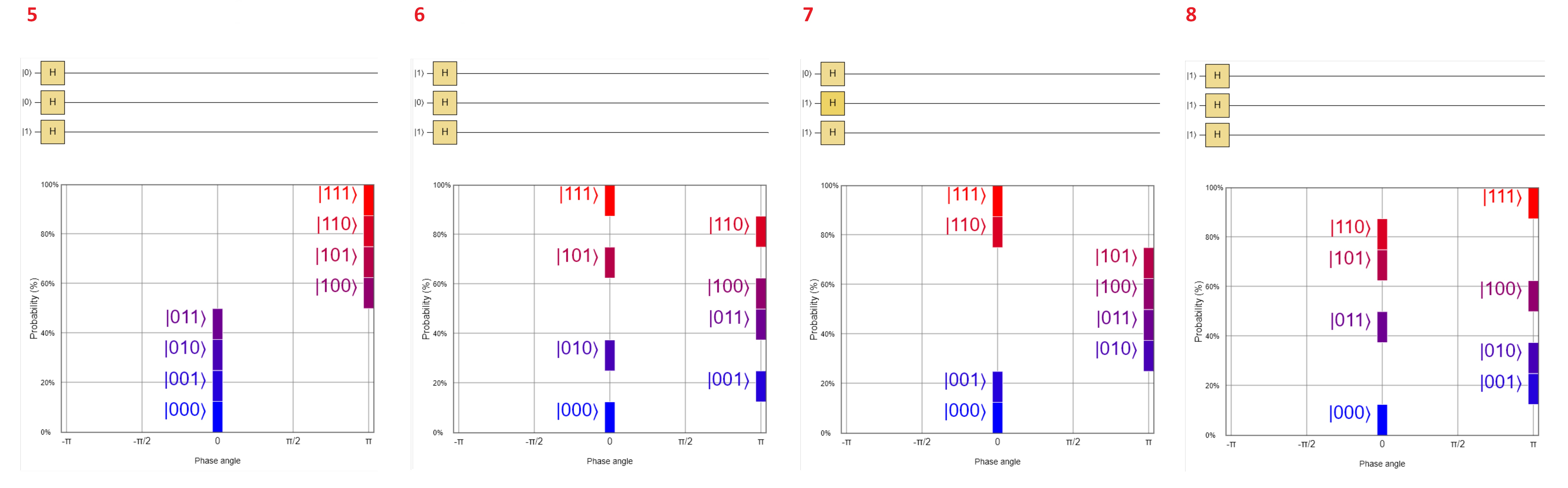}}
    }
}
\caption{\label{fig:Hadamard 3q}Hadamard transformation for all $8$ classical values of a 3-\qbit{} register.}
\end{figure}

Next we look at a \multiqbit{} system. We take the example of a $3$-\qbit{} system and initialize the 3 \qbits{} with classical values: $\ket{x_0}, \ket{x_1}, \ket{x_2}, x_i \in \{0, 1\}$. Their tensor product is a state of the computational basis $\ket{x_2 x_1 x_0}$. After applying a Hadamard gate operation to each \qbit{} we get a superposition state $\frac{1}{\sqrt{2}}\ket{0}\pm\frac{1}{\sqrt{2}}\ket{1}$. The sign in front of $\ket{1}$ depends on the original state of the \qbit{} and is positive for $\ket{0}$ and negative for $\ket{1}$. In total the tensor product of these superposition states results in the sum of all $8$ states of the computational basis with equal magnitude $\frac{1}{\sqrt{2^3}}$ and a sign that depends on how many of the $\ket{1}$ stem from a $\ket{1}$ before the Hadamard operation; if this number is even we have a positive sign, if it is odd the sign is negative. Especially the transformation of $\ket{000}$ has only positive phases. Figure \ref{fig:Hadamard 3q} shows for all $8$ states of the computational basis the Hadamard transformation.

\subsection{Visualization of the Deutsch-Josza Algorithm }
The Deutsch-Josza algorithm \cite{Deutsch1992} was the first algorithm that demonstrated a proven exponential advantage for a quantum algorithm compared to equivalent classical algorithms. For a given function $f:\mathbb{Z}_{2^n} \to \mathbb{Z}_2$ it is known, that it is constant ($f(x)$ is the same for all $x$) or balanced (the number of inputs $x$ for which $f(x)=0$ equals the number of inputs $x$ for which $f(x)=1$, i.e. $|f^{-1}[\{0\}]|=f^{-1}[\{1\}]|$). For a classical algorithm the function has to be tested for different arguments. Worst case $2^{n-1}$ tested arguments return the same value, then it is still unclear if the function is constant or balanced; it needs an additional test. If $f$ returns for the $2^{n-1}+1^{st}$ argument the same value again, then the function cannot be balanced anymore, and therefore $f$ must be constant. If the value is different from the previous tests, then the function is not constant and therefore must be balanced. We need $2^{n-1}+1$ test rounds in the worst case, so the complexity of the classical algorithm is $O(2^n)$ in $n$, where $n$ is the number of bits in the input.

In the next two paragraphs we will analyze the circuit of the Deutsch-Jozsa algorithm, which can decide after a single  round if the given function is constant or balanced. That means the complexity of the quantum algorithm is $O(1)$. We will use the \Stateogram{} visualization. We will fix our dimension to $n=2$, nevertheless the argumentation and the \Stateogram{} visualization itself remain valid for any problem size. During the algorithm the states of the computational basis occur multiplied with probability amplitudes with phase angles of $0$ and $\pi$; for better readability we will not write the magnitude of the probability amplitudes but only the sign, so we write $+1$ or $-1$ as shorthand for the amplitudes.

\subsubsection{Constant Functions}
\begin{figure}
\centering
\fbox{\includegraphics[width=0.98\linewidth]{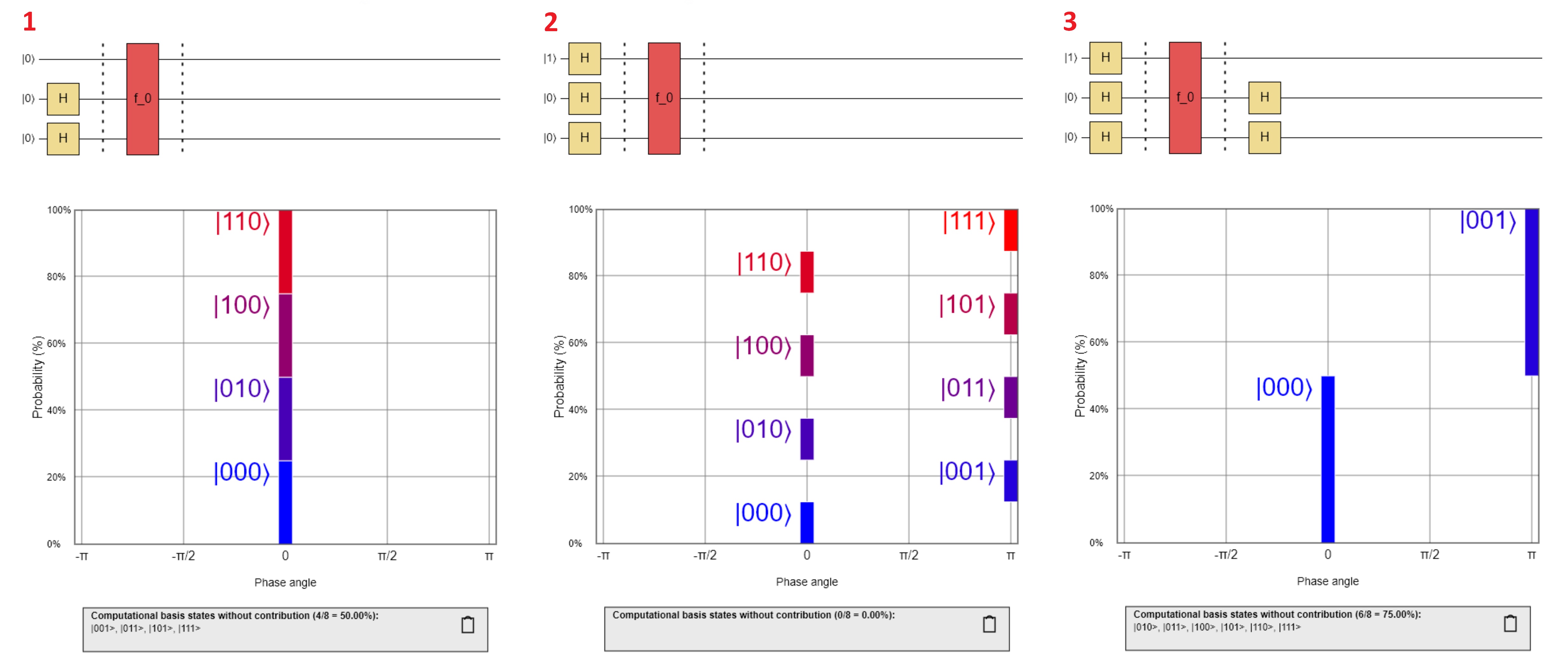}}
\caption{\label{fig:dj_const 3q}Deutsch-Josza for constant function}
\end{figure}

The secret function is implemented as a circuit on three \qbits{} $\ket{x_2 x_1 x_0}$. The argument bits $\ket{x_1}$ and $\ket{x_2}$ are not changed by the circuits. Bit $\ket{x_0}$ is for the calculated function value and the circuit changes this by returning $x_0 \oplus f(x_2 2^1 + x_1 2^0)$. We start with a constant function $f_0$ that maps all arguments to $0$. Figure \ref{fig:dj_const 3q} starts under (1) with all three \qbits{} initialized to $\ket0$. The Hadamard gates on $\ket{x_1}$ and $\ket{x_2}$ split the values of those \qbits{} to $\ket{0}$ and $\ket{1}$ thereby creating a superposition of all combinations of $\ket{x_2, x_1}$. The circuit calculates the values of all these combinations. Since the function is constant $0$, it leaves $\ket{x_0}$ as $\ket{0}$. So the \Stateogram{} shows the value table of $f$. Now lets start again with a slightly different initialization: we insert a Hadamard gate on $\ket{x_0}$ as well and initialize $\ket{x_0} = \ket{1}$. This splits $\ket{x_0}$ into $\ket{0}$ and $-\ket{1}$ as shown in figure \ref{fig:dj_const 3q}(2). The constant function output of $0$ is added modulo 2 to $\ket{0}$ and $-\ket{1}$. Therefore we get the value table of $f$ at phase angle $\phi = 0$ and a value table of function $1 \oplus f$ with phase angle $\phi = \pi$. Finally we apply Hadamard gates to $\ket{x_1}$ and $\ket{x_2}$. Since we have all combinations $\ket{x_2, x_1}$ multiplied with $\ket{x_0}=\ket{0}$ the Hadamard operation will join these into $\ket{x_2 x_1 x_0} = \ket{000}$. Accordingly all combinations $\ket{x_2, x_1}$ multiplied with $\ket{x_0}=-\ket{1}$ are joined into $\ket{x_2 x_1 x_0} = -\ket{001}$ as shown in figure \ref{fig:dj_const 3q}(3). Therefore a measurement of $\ket{x_2, x_1}$ will always return $\ket{00}$. The same argument can be done for the constant function $f_1$, which maps all inputs to $1$.

\subsubsection{Balanced Functions}
\begin{figure}
\centering
\fbox{\includegraphics[width=0.98\linewidth]{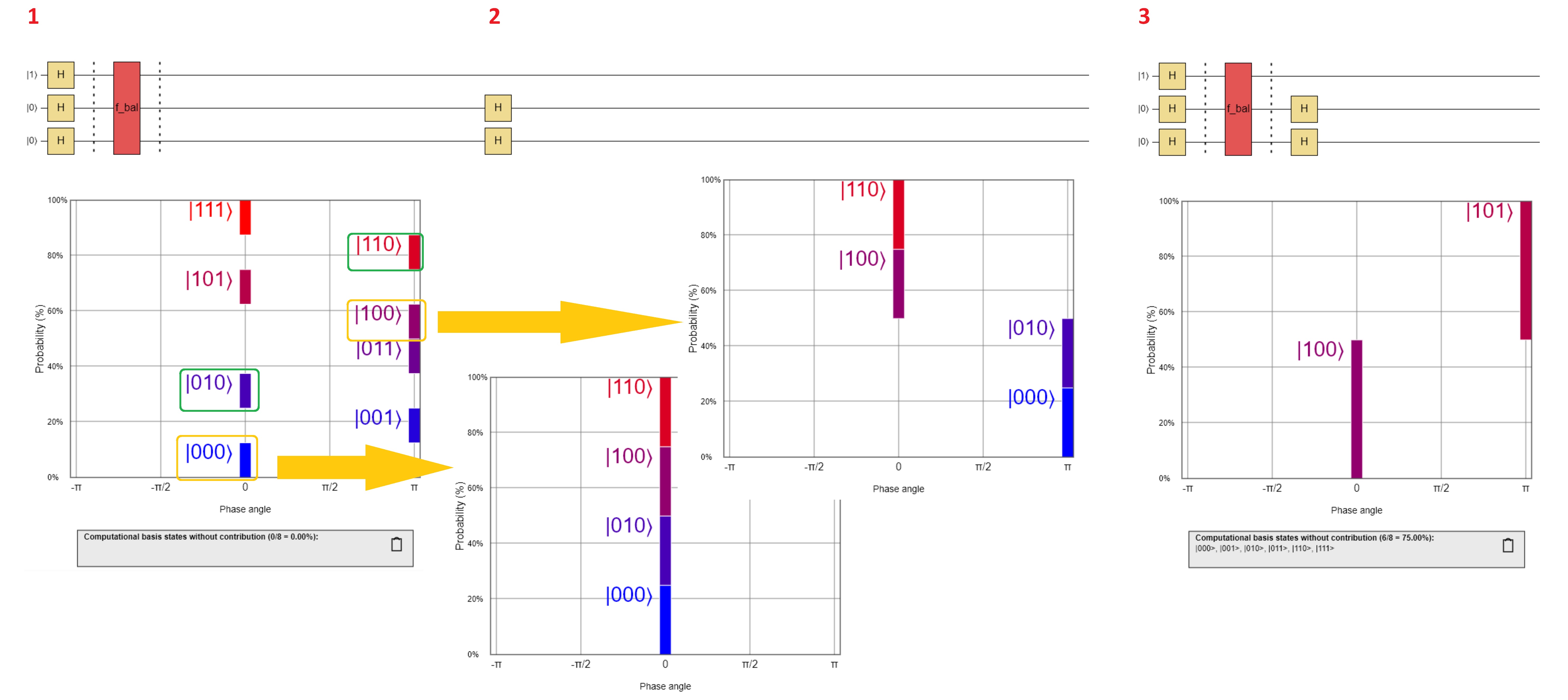}}
\caption{\label{fig:dj_bal3q}Deutsch-Josza for balanced function}
\end{figure}

Finally we analyze the Deutsch-Jozsa algorithm for the case of a balanced function $f_{bal}$. Initializing $\ket{x_2 x_1 x_0} =\ket{001}$ and applying a Hadamard transformation and the circuit for the function $f_{bal}$ is shown in figure \ref{fig:dj_bal3q}(1). Similar to figure \ref{fig:dj_const 3q}(2) we get the value table of $f_{bal}$ at phase angle $\phi = 0$ and a value table of function $1 \oplus f_{bal}$ at phase angle $\phi = \pi$. 
First we focus on all basis states with $\ket{x_0}=\ket{0}$. Since $f$ and $1 \oplus f$ are balanced functions half of the output values of each function are $0$. Therefore we can map one-to-one the basis vectors of form $\ket{x_2 x_1 0}$ with phase angle $\phi=0$ to those of form $\ket{x_2 x_1 0}$ with phase angle $\phi=\pi$. Let us now examine how the two members of such a pair are transformed by the Hadamard gates applied to \qbits{} $\ket{x_2}$ and $\ket{x_1}$. In figure \ref{fig:dj_bal3q}(2) the transformation for the pair marked by yellow boxes is shown. The basis state $\ket{0 0 0}$ is transformed into 4 equally probable state vectors where $\ket{0 0 0}$ as part of the resulting state, has a positive sign (phase angle $\phi=0$). The partner in the pair is $-\ket{1 0 0}$, which is transformed into 4 equally probable state vectors. Since we have a basis state with a negative amplitude all components of the resulting states have to be multiplied by $-1$. Thus, the amplitude of $\ket{000}$ in the resulting state receives a positive contribution from the Hadamard transformation and a negative contribution (due to the initial negative coefficient of the basis state). Consequently, the contributions to $\ket{000}$ from the Hadamard transformation of these two basis states cancel each other out due to destructive interference, resulting in a zero amplitude for $\ket{0 0 0}$. The same argument is valid for all other pairs and therefore the amplitude of the basis state $\ket{0 0 0}$ vanishes. 
Next we focus on all basis states with a $\ket{x_0} = \ket{1}$. With the same argument we conclude, that the basis state $\ket{0 0 1}$ vanishes after the Hadamard transformation. 
In both cases a measurement of $\ket{x_2 x_1}$ will never return $\ket{00}$. 

\subsubsection{Measurement}
In the previous two sections we observed, that a measurement of $\ket{x_2, x_1}$ will always return $\ket{00}$ for a constant function and never return $\ket{00}$ for a balanced function. So we can decide by a single measurement if the function included in the respective Deutsch-Josza circuit is a constant or balanced function. This proves the exponential advantage of the quantum algorithm compared to classical implementations.

\section{Discussion and Future Work}
We introduced a novel visualization for the superposition states of \qbits{} and \qregs{}. In several examples for elementary gate operations like Hadamard gate and for more complex circuits like Deutsch-Josza we evaluated the visualization. It was demonstrated how the insights given by the visualization can help understanding and verifying quantum algorithms. To facilitate rapid prototyping and testing the \Stateogram{} has been integrated into a version of the open source quantum computing simulator Quirk \cite{Qui14} or Quirk-E \cite{Qui25} respectively. The integrated version can be found as quirk-s \cite{Quis25}. With this tool every step of a quantum algorithm can be tracked and the respective intermediate state of the quantum computer is displayed as a comprehensive \Stateogram{} image. We are planning to provide integrations of the \Stateogram{} visualization with other open source frameworks for quantum computing such as the Qulacs package.

\section{Conclusion}
The \Stateogram{} delivers a novel comprehensive visualization of the state inside a quantum computer during the execution of a quantum algorithm. The visualization can be beneficial to explain non-experts complex phenomena of quantum algorithms and some of the computational mechanisms inside quantum computers. 
The visual calculus provided by the \Stateogram{} can directly illustrate the effect of operations on computational data. So it can help experienced programmers as well to draft, explain and test new quantum gate algorithms.

\textit{Acknowledgements} -- I would like to thank Stefan Walter, Christian Münch, Andreas Rohnfelder and Markus Kirsch for their stimulating discussions, reviews and valuable comments by . Special thanks are extended to Craig Gidney and the Finnish team (Samuel Ovaskainen, Majid Haghparast, Ronja Heikkinen, and Julian Fuchs) for developing the excellent open-source quantum computer simulators Quirk and Quirk-E, which provided a perfect environment for state visualization.

\printbibliography

@article{Bloch1946,
  author={Bloch, F.},
  journal={Physical Review},
  title={Nuclear Induction},
  year={1946},
  volume={70},
  number={7-8},
  pages={460},
  doi={10.1103/PhysRev.70.460}
}

@article{Arecchi1972,
  title = {Atomic Coherent States in Quantum Optics},
  author = {Arecchi, F. T. and Courtens, Eric and Gilmore, Robert and Thomas, Harry},
  journal = {Phys. Rev. A},
  volume = {6},
  issue = {6},
  pages = {2211--2237},
  numpages = {0},
  year = {1972},
  month = {12},
  publisher = {American Physical Society},
  doi = {10.1103/PhysRevA.6.2211},
}

@article{Deutsch1992,
    author = {Deutsch, David  and Jozsa, Richard },
    title = {Rapid solution of problems by quantum computation},
    journal = {Proceedings of the Royal Society of London. Series A: Mathematical and Physical Sciences},
    volume = {439},
    number = {1907},
    pages = {553-558},
    year = {1992},
    doi = {10.1098/rspa.1992.0167}
}

@misc{ibm14,
  author={{IBM Quantum}},
  title={{IBM Qiskit}},
  howpublished={\url{https://www.ibm.com/quantum/qiskit}},
  note={Accessed: 2025-08-01}
}

@article{BSR21,
  author={Bardin, J. C. and Slichter, D. H. and Reilly, D. J.},
  journal={IEEE Journal of Microwaves},
  title={Microwaves in Quantum Computing},
  year={2021},
  volume={1},
  number={1},
  pages={403-427},
  doi={10.1109/JMW.2021.3059438}
}

@phdthesis{Wil21,
  author={Williams, M. M.},
  title={{QCVis: A quantum circuit visualization and education platform for novices}},
  school={Harvard University},
  year={2021}
}

@article{Ruan2023,
  author={Ruan, Shaolun and Yuan, Ribo and Guan, Qiang and Lin, Yanna and Mao, Ying and Jiang, Weiwen and Wang, Zhepeng and Xu, Wei and Wang, Yong},
  title={{VENUS: A Geometrical Representation for Quantum State Visualization}},
  journal={COMPUTER GRAPHICS forum},
  volume={42},
  number={3},
  pages={247-258},
  year={2023},
  publisher={Wiley Online Library},
  doi={10.1111/cgf.14827} 
}

@inproceedings{AJMK09,
  author={Altepeter, J. B. and Jeffrey, E. R. and Medic, M. and Kumar, P.},
  booktitle={2009 Conference on Lasers and Electro-Optics and 2009 Conference on Quantum Electronics and Laser Science Conference},
  title={Multiple-qubit quantum state visualization},
  year={2009},
  volume={},
  number={},
  pages={1-2},
  doi={10.1364/CLEO.2009.JThA10}
}

@inproceedings{HD04,
  author={Havel, T. F. and Doran, C. J.},
  booktitle={Quantum Information and Computation II},
  title={A bloch-sphere-type model for two qubits in the geometric algebra of a 6d euclidean vector space},
  year={2004},
  volume={5436},
  pages={93-106},
  organization={SPIE}
}

@article{GI12,
  author={Galambos, M. and Imre, S.},
  journal={International Journal on Advances in Systems and Measurements},
  title={Visualizing the effects of measurements and logic gates on multi-qubit systems using fractal representation},
  year={2012},
  volume={5},
  number={1 \& 2}
}

@article{CMM17,
  author={Chernega, V. N. and Man'ko, O. V. and Man'ko, V. I.},
  journal={Journal of Russian Laser Research},
  title={Triangle geometry of the qubit state in the probability representation expressed in terms of the triada of malevich's squares},
  year={2017},
  volume={38},
  number={2},
  pages={141-149}
}

@inproceedings{PRZ17,
  author={Paykin, J. and Rand, R. and Zdancewic, S.},
  booktitle={Proceedings of the ACM SIGPLAN International Conference on Functional Programming},
  title={{Qwire: A core language for quantum circuits}},
  year={2017},
  pages={846--858}
}

@article{ZW18,
  author={Zulehner, A. and Wille, R.},
  journal={IEEE Transactions on Computer-Aided Design of Integrated Circuits and Systems},
  title={{Advanced simulation of quantum computations}},
  year={2018},
  volume={38},
  number={5},
  pages={848-859},
  doi={10.1109/TCAD.2018.2862809}
}

@article{RWJ2022,
  author={Ruan, Shaolun and Wang, Yong and Jiang, Weiwen and Mao, Ying and Guan, Qiang},
  title={{Vacsen: A visualization approach for noise awareness in quantum computing}},
  journal={arXiv preprint arXiv:2207.14135},
  year={2022}
}

@inproceedings{TPCW17,
  author={Tao, Z. and Pan, Y. and Chen, A. and Wang, L.},
  booktitle={2017 International Conference on Virtual Reality and Visualization (ICVRV)},
  title={{Shorvis: A comprehensive case study of quantum computing visualization}},
  year={2017},
  pages={360-365},
  doi={10.1109/ICVRV.2017.8285987}
}

@misc{Qui14,
  author={{Gidney, Craig (Strilanc), }},
  title={{Quirk}},
  howpublished={\url{https://github.com/Strilanc/Quirk}},
  note={Accessed: 2025-08-21}
}

@misc{Qui25,
  author={{Ovaskainen, Samuel and Haghparast, Majid and Heikkinen, Ronja and Fuchs, Julian, University of Jyväskylä}},
  title={{Quirk-E}},
  howpublished={\url{https://github.com/DEQSE-Project/Quirk-E}},
  note={Accessed: 2025-08-21}
}

@misc{Quis25,
  author={{Schinkel, Fritz}},
  title={{quirk-s}},
  howpublished={\url{https://github.com/fritz-schinkel/quirk-s}},
  note={Accessed: 2025-08-21},
  year={2025}
}

\end{document}